\documentclass[12pt,thmsa]{article}
\usepackage{amssymb}
\begin{document}

\begin{center}
{\LARGE Quotient Construction of 't Hooft's Quantum Equivalence Classes }

\medskip

C.P.Sun$^a,$X.F.Liu $^{a,b}$and S.X.Yu$^a$

$^a$Institute of Theoretical Physics, Chinese Academy of Sciences,

Beijing 100080, China
\end{center}

\begin{quotation}
$^b$Department of Mathematics, Peking University, Beijing, 100871,China
\end{quotation}

\begin{center}
\bigskip

\textbf{Abstract}
\end{center}

\begin{quotation}
\textit{\ Most recently 't Hooft has postulated (\textrm{G 't Hooft, Class.
Quant. Grav. 16 (1999) 3263-3279}) that quantum states at the ``atomic
scale''can be understood as equivalence classes of primordial states
governed by a dissipative deterministic theory underlying quantum theory at
the ``Planck scale''. Defining invariant subspaces clearly for primordial
states according to a given evolution, we mathematically re-formulate 't
Hooft's theory as a quotient space construction with the time-reversible
evolution operator induced naturally. With this observation and some
analysis , 't Hooft's theory is generalized beyond his case where the
evolution at the ``Planck scale'' is periodic or the time is discrete. We
also give a novel illustration that the Fock space of quantum oscillator
could follow from the quotient space construction for certain primordial
states obeying non-reversible evolution governed by a non-Hermitian
Hamiltonian.}
\end{quotation}

\section{\protect\bigskip Introduction}

To probe the physical differences in locality and causality between the so
called Planck scale physics such as quantum gravity and the usual quantum
field theories in some flat background space-time, Gerard't Hooft postulated
[1,2] that there should be a dissipative deterministic theory underlying the
usual quantum theory. In his theory, the generic quantum mechanics is no
longer the crucial starting point. Rather, a deterministic theory with
dissipation of information at the Planck scale is needed to derive quantum
mechanics at the atomic scale. Quantum state used to make probabilistic
prediction about physical phenomenon is then shown to be a derived concept.

In 't Hooft's opinion, at the atomic scale quantum states are equivalence
classes of primordial states at the Planck scale. If we only care the
temporal evolution of equivalence classes, the information within each
equivalence class can be ignored. Then from a non-time-reversible evolution,
which characterizes a deterministic process with dissipation at the Planck
scale, we can obtain a time-reversible evolution of the properly defined
equivalence classes for primordial states. Taking the equivalence classes to
be quantum states we are then able to introduce a unitary evolution law at
the atomic scale. Apparently, here the central problem is how to classify
the Planck scale states with respect to a deterministic evolution.'t Hooft's
solution to this problem is as follows. He argued that two Planck scale
states are equivalent at the atomic scale if , after some finite time
interval, they evolve into the same state. This leads to a natural
definition of equivalence classes: two states are in the same equivalence
class if and only if they evolve into the same state after some finite time
interval. Quantum states are identified with these equivalence classes.

To see 't Hooft's idea clearly,we will make use of mathematical
terminologies such as quotient space and induced representation of
operators. We will first properly define an invariant subspace of primordial
states related to the equivalence classes defined by 't Hooft. Then we can
identify the space of quantum states,which is spanned by the equivalence
classes according to 't Hooft, with the quotient space and naturally
re-formulate the time-reversible evolution at the atomic scale by the
mechanism of induced representation of the dissipative deterministic
evolution operator on the quotient space. Finally, we extend 't Hooft's
theory to cases where the evolution of primordial states is not necessarily
periodic at the Planck scale( 't Hooft has implicitly assumed the cyclic
evolution law in the case of discrete time variable). Based on our
generalization we try to understand the quantum oscillator and its Fock
space from the quotient space construction for certain primordial states
obeying a non-reversible evolution governed by a non-Hermitian Hamiltonian.

\section{Quotient Representation of quantum states}

In 't Hooft's theory[1], primordial states at the Planck scale need not form
a linear space. Generally they can be denoted by a set $\Sigma $ $=\{\phi
_i|i\in I\}$, where $I$ stands for an index set. The underlying
deterministic evolution is a transformation $U$ (usually depending on time)
of $\Sigma $ to itself. It can be represented by a matrix with the entries $%
0 $ or $1$ if $I$ is a countable set. The determinism requires that there be
at most one nonzero entry in each column. Otherwise , the system will be
forced to evolve into an uncertain state, namely, a superposition of several
elements that is not in $\Sigma .$ As $U$ is an evolution operator, we write
it as $U=U(t_f,t_{i,})$ by convention. Physically, it represents the
evolution in the time interval $[t_i,t_f].$ Certainly the evolution should
satisfy the so called semi-group condition
\[
U(t_f,t_m)U(t_m,t_i)=U(t_f,t_i)
\]
\begin{equation}
U(t,t)=1
\end{equation}
In general, $U$ is singular, namely, it has no inverse.Such singular
operator describes deterministic process with dissipation. As a matter of
fact, under such an evolution some states will disappear and some states
will evolve into the same state, or in other words, some states with a
different past may have the same deterministic fate. 't Hooft thinks that,
if two states evolve in such a way that their futures are identical, they
should represent the same state at the atomic scale. In this view, he
divides the elements of $\Sigma $ into equivalence classes,$\phi _{i_1}$ and
$\phi _{i_2}$ ($i_{1,}i_2\in I$) being in the same equivalence class if they
are evolved into the same state after finite time interval. Denote by $\Xi
=\{\overline{\phi }_j|j\in J\}$ the set of the equivalence classes, where $J$
is another index set.Then 't Hooft postulates that the space of quantum
states is spanned by $\{\overline{\phi }_j|j\in J\}$ and claims that the
reduced evolution on the space of quantum states is reversible. Now let us
analyze 't Hooft's theory from mathematical point of view as follows.

Let $V$ be the vector space spanned by $\{\phi _i|i\in I\}.$Then $%
U(t_f,t_{i,})$ can be extended to a linear transformation of $V$. We will
call $V$ the space of primordial states in spite of the fact that generally
it contains elements which are not primordial states. Let $V_1$ denote the
subspace of $V$ consisting of the vectors annihilated by $U(0,t)$ at some $t$%
, namely, a vector $v$ belongs to $V_1$if and only if there exists some $%
U(t,0)$ such that $U(t,0)v=0$. Now it is easy to observe that \textit{The
space of quantum states is none other than the quotient space }
\begin{equation}
V/V_1=\{|\phi \rangle \triangleq \phi +V_1|\phi \in V\}.
\end{equation}
It is also easy to notice that 't Hooft's construction implies the
assumption that the evolution operator $U(t_2,t_1)$ only depend on the
difference of $t_2$and $t_1$, i.e., we can write $U(t_2,t_1)=U(t_2-t_1)$.
Indeed , if this is the case, a non-singular evolution law of the quantum
states naturally follows from $U(t_2,t_1).$ Otherwise, generally the
evolution operator at the Planck scale cannot be reduced to the space of
quantum states at the atomic space. Mathematically, this is because, for a
linear transformation in $End(V)$ to have an induced action on the quotient
space $V/V_1$ , $V_1$ should be invariant with respect to it. Let $\overline{%
v}$ $\equiv |\nu \rangle $ denote the equivalence class containing $v$ . We
notice that $V_1$is invariant under $U(t_2,t_1)$ in this case. Thus $%
U(t_2,t_1)$ induces a natural action on the quotient space $V/V_1.$ We
denote the induced operator by $\overline{U(t_2,t_1)},$then we have
\begin{equation}
\overline{U(t_2,t_1)}\overline{v}=\overline{U(t_2,t_1)v}.
\end{equation}
It is easily seen that $\overline{U(t_2,t_1)}$ is non-singular, i.e., zero
is not its eigenvalue. In fact, if $\overline{U(t_2,t_1)}\overline{v}=%
\overline{0}$ , then $U(t_2,t_1)v\in V_1.$Thus there exists some $t$ such
that $U(t,0)U(t_2,t_1)v=0.$ It then follows that
\[
U(t,0)U(t_2,t_1)v=U(t_2+t,t_2)U(t_2,t_1)v
\]
\begin{equation}
=U(t_2+t,t_1)v=U(t_2-t_1+t,0)v=0.
\end{equation}
By definition this means $v\in V_1,$i.e.,$\overline{v}=\overline{0}$.This
proves the non-singularity of $\overline{U(t_2,t_1)}.$ Now the unitarity of $%
\overline{U(t_2,t_1)}$ remains to be established. We will handle this
problem in a special case below.

If the condition $U(t_2,t_1)=U(t_2-t_1)$ is not satisfied, to guarantee the
non-singularity of $\overline{U(t_2,t_1)}$ , the definition of the invariant
subspace $V_1$ needs to be modified. It seems that we should define $V_1$ in
the following way:

\begin{quotation}
\textit{A vector }$v$\textit{\ belongs to }$V_1$\textit{if and only if there
exist finitely many }$t_i(i=1,2,\cdots ,r)$\textit{\ such that }$%
U(t_1,t_2)U(t_3,t_4)\cdots U(t_{r-1},t_r)v=0.$
\end{quotation}

Unfortunately, in this definition the physical meaning of $V_1$ is unclear.
Let us return to the case with the condition $U(t_2,t_1)=U(t_2-t_1)$.
Following 't Hooft, we consider a system with discrete time coordinates. We
assume that the time $t$ takes values in $Z^{+}$, the set of non-negative
integers. Actually, the system is periodic since we have
\begin{equation}
U(n+1,n)=U(1,0),U(n,0)=U(1,0)^n
\end{equation}
for $n\in Z^{+}$, and the invariant subspace $V_1$ is
\begin{equation}
V_1=\{v\in V|\exists n\in Z^{+}\ s.t.\ U(1,0)^nv=0\}
\end{equation}
In Ref.[1,2] 't Hooft presented a simple example to illustrate his theory.
Fit into the above mathematical framework, The example goes as follows: $V$
is four dimensional:
\begin{equation}
V=span\{v_1,v_2,v_3,v_4\}
\end{equation}
and
\[
U(1,0)=\left(
\begin{array}{cccc}
0 & 0 & 1 & 0 \\
1 & 0 & 0 & 1 \\
0 & 1 & 0 & 0 \\
0 & 0 & 0 & 0
\end{array}
\right)
\]
with respect to the basis $\{v_1,v_2,v_3,v_4\}$.It is easily seen that
\[
V_1=span\{v_1-v_4\}
\]
\[
V/V_1=span\{\overline{v}_1,\overline{v}_2,\overline{v}_3\}
\]
and the induced evolution operator is
\[
\overline{U(1,0)}=\left(
\begin{array}{ccc}
0 & 0 & 1 \\
1 & 0 & 0 \\
0 & 1 & 0
\end{array}
\right)
\]
with respect to the basis $\{\overline{v}_1,\overline{v}_2,\overline{v}_3\}$%
. Clearly,$\overline{U(1,0)}$ is unitary relative to a properly defined
inner product. This is not at all accidental. In fact, \textit{if the space
of primordial states is finite dimensional a dissipative deterministic
evolution at the Planck scale always induces a unitary evolution at the
atomic scale if only we choose an inner product on the space of quantum
states adequately.}We will present the proof elsewhere.

\section{Non-periodic Dynamics}

We now turn to consider non-periodic evolution process, such as scattering
process, with time variable approaching infinity. Assume such a process is
described by an evolution operator $U(0,+\infty )\triangleq W$ at the Planck
scale. As above, let $V$ be the space of primordial states. Suppose $V$ is
finite dimensional. Inspired by 't Hooft's theory, we postulate that the
space of quantum states at the atomic scale is the quotient space $V/V_1$
with $V_1$ defined as follows:
\[
V_1=\{v\in V|\exists n\in Z^{+}\ s.t.\ W^nv=0\}.
\]
Suppose that the characteristic polynomial of $W$ is
\begin{equation}
p_w(\lambda )=\prod\limits_{i=0}^r(\lambda -\lambda _i)^{m_i}
\end{equation}
where $\lambda _0=0$ and $\lambda _j\neq 0$ for $j\neq 0.$ Obviously, $V_1$
is just the kernel of $W^{m_0},$namely,
\begin{equation}
V_1=KerW^{m_0}=\{v\in V|\ W^{m_0}v=0\}
\end{equation}
and the characteristic polynomial of the induced operator $\overline{W}$ is
\begin{equation}
p_{\overline{w}}(\lambda )=\prod\limits_{i=1}^r(\lambda -\lambda _i)^{m_i}.
\end{equation}
Therefore, $\overline{W}\in End(V/V_1)$ is non-singular. Let us go on to
deal with the unitarity problem of $\overline{W}.$

$\overline{W}$ is called\textit{\ unitarizable }if it is diagonalizable and
all of its eigenvalues are of modulus $1$ . By definition, if $\overline{W}$
is unitarizable, there exists a basis $\{\overline{v}_1,\overline{v}%
_2,\cdots ,\overline{v}_m\}$ of $V/V_1$ such that $\overline{W}\overline{v}%
_j=e^{i\theta _j}\overline{v}_j\ (j=1,2,\cdots ,m)$ where $\theta _j$ is a
real number. Therefore, if we define the ``canonical'' inner product $(\ ,\ )
$ on $V/V_1$ satisfying $(\overline{v}_i,\overline{v}_j)=\delta _{ij}$, then
$\overline{W}$ is unitary with respect to it. We have shown that if an
operator is unitarizable it can be made unitary by properly defining an
inner product on the space that it acts on, as the term suggests. The
converse statement is trivially true, as one easily sees.As for the
unitarizability condition for $\overline{W}$,it is not difficult to show
that $\overline{W}$ is unitarizable if and only if the minimal polynomial of
$W$ is of the form $p(\lambda)=\lambda^n\prod_{j=1}^{m}(\lambda
-e^{i\theta_j})$ where $\theta_j\ (j=1,2,\cdots,m)$ are different nonzero
real numbers.

If $\overline{W}$ is not unitarizable we can construct a unitary operator
from $\overline{W}$ by the polar decomposition of $\overline{W}.$
Explicitly, we define
\begin{equation}
\overline{U}_{\overline{w}}=\overline{W}(\overline{W}^{+}\overline{W}%
)^{-\frac 12}
\end{equation}
It is then elementary to show the unitarity of $\overline{U}_{\overline{w}}.$
Certainly, $\overline{U}_{\overline{w}}$ depends on the inner product on $%
V/V_1$. But it is always unitary with respect to the chosen inner product.
It is also clear that when $\overline{W}$ is unitarizable $\overline{U}_{%
\overline{w}}$ coincides with $\overline{W}$ if we choose the ``canonical''
inner product on $V/V_1.$In general, there does not exist a canonical way to
construct a unitary operator from $\overline{W}$. This corresponds to the
fact that there does not exists a canonical way to introduce an inner
product on $V/V_1.$

We proceed to present a method of obtaining the matrix representation of $%
\overline{W}.$ Denote by $(KerW^{m_0})^{\bot }$ the orthogonal complement to
the subspace $KerW^{m_0}$ in $V$ with respect to the inner product on $V.$
Then we have the decomposition
\[
V=KerW^{m_0}\oplus (KerW^{m_0})^{\bot }.
\]
Obviously, the operator $W^{+m_0}W^{m_0}$ is Hermitian and $%
(KerW^{m_0})^{\bot }$ is a $W^{+m_0}W^{m_0}$ -invariant subspace. Thus the
restriction of $W^{+m_0}W^{m_0}$ to $(KerW^{m_0})^{\bot }$ is also
Hermitian. Hence there are eigenvectors $v_1,v_2,\cdots ,v_d$ of $%
W^{+m_0}W^{m_0}$ such that they constitute a basis of $(KerW^{m_0})^{\bot }$%
. Choose a basis $\{v_{d+1},v_{d+2},\cdots ,v_N\}$ of $KerW^{m_0}.$Then $%
\{v_i|i=1,2,\cdots ,N\}$ is a basis of $V$ and $\{\overline{v}_i\triangleq
\overline{v}_i+V_1|i=1,2,\cdots ,d\}$ is a basis of $V/V_1.$Suppose $P$ is
the projection operator upon $(KerW^{m_0})^{\bot }.$Then the operator $PWP$
has the following matrix representation
\begin{equation}
PWP=\left(
\begin{array}{cc}
M & 0 \\
0 & 0
\end{array}
\right)
\end{equation}
with respect to the basis $\{v_i|i=1,2,\cdots ,N\},$where $M$ is the matrix
representation of $\overline{W}$ with respect to the basis $\{\overline{v}%
_i|i=1,2,\cdots ,d\}.$

To illustrate the above arguments,let us take

\[
W=\left(
\begin{array}{llll}
0 & 0 & 1 & 0 \\
1 & 1 & 0 & 0 \\
0 & 0 & 0 & 0 \\
0 & 0 & 0 & 1
\end{array}
\right)
\]
as an example.Its characteristic polynomial is $\left( \lambda -1\right)
^2\lambda ^2.$ Thus the invariant subspace $V_1=KerW^2.$ By simple
calculation we have
\[
V_1^{\perp }=span\{v_1,v_2\},\ \ V_1=span\{v_3,v_4\}
\]
where
\[
v_1=(0,0,0,1)^T,v_2=(1,1,1,0)^T,v_3=(1,0,-1,0)^T,v_4=(0,1,-1,0)^T.
\]
Then we obtain the matrix representation of $\overline{W}$ relative to the
basis $\{\overline{v}_1,\overline{v}_2\}:$%
\[
M=\left(
\begin{array}{ll}
1 & 0 \\
0 & 1
\end{array}
\right) .
\]

To sum up, in the finite dimensional case, if we can find an adequate
classification of the so called primordial states at the Planck scale such
that the information concerning the time irreversibility could be reasonably
ignored , then at the atomic scale we can manage to obtain a unitary
evolution, describing a quantum mechanical process. When we pass to the
infinite dimensional case the situation becomes subtle and hard to manage.
Especially, it should be difficult to identify the subspace $V_1$.
Nevertheless, the central idea of classifying primordial states and
identifying quantum states with equivalence classes is applicable without
difficulty. 't Hooft has shown us two elegant examples: the classical motion
with limit cycles and the massless neutrinos moving as a plane in space-time
[1,2]. We wish to try our hands at discrete infinite dimensional case in the
next section.

\section{ Fock states as equivalence classes}

It is well known that Fock space is a central concept in quantum field
theory. As an introductory example, let us analyze this concept from the
above elucidated viewpoint. We recall that the rank one Heisenberg-Wely
algebra $\mathcal{W}_1$ is a Lie algebra generated by the elements $A,B,%
\mathbf{1}$ with the commutation relation $[A,B]=\mathbf{1}$ where $\mathbf{1%
}$ is the central element. The one particle Fock space is an infinite
dimensional irreducible representation of $\mathcal{W}_1$. Mathematically, $%
\mathcal{W}_1$ has an intrinsic representation space $V$ constructed from
its universal enveloping algebra[4]: $V$ is a vector space with a basis
\[
\{f(m,n)\triangleq A^mB^n\Omega |m,n\in Z^{+}\};
\]
$A,B$ acts on $V$ naturally and $\mathbf{1}\Omega =\Omega $. $f(m,n)$ might
be understood as a primordial states at the Planck scale and they span the
``Planck Space'' . By definition we have
\[
Af(m,n)=f(m+1,n)
\]
\begin{equation}
Bf(m,n)=f(m,n+1)+mf(m-1,n)
\end{equation}
The actions of $A$ and $B$ are illustrated in Fig.1a where
$f(m,n)$ is denoted by a point $(m,n)$.
\begin{figure}
\include{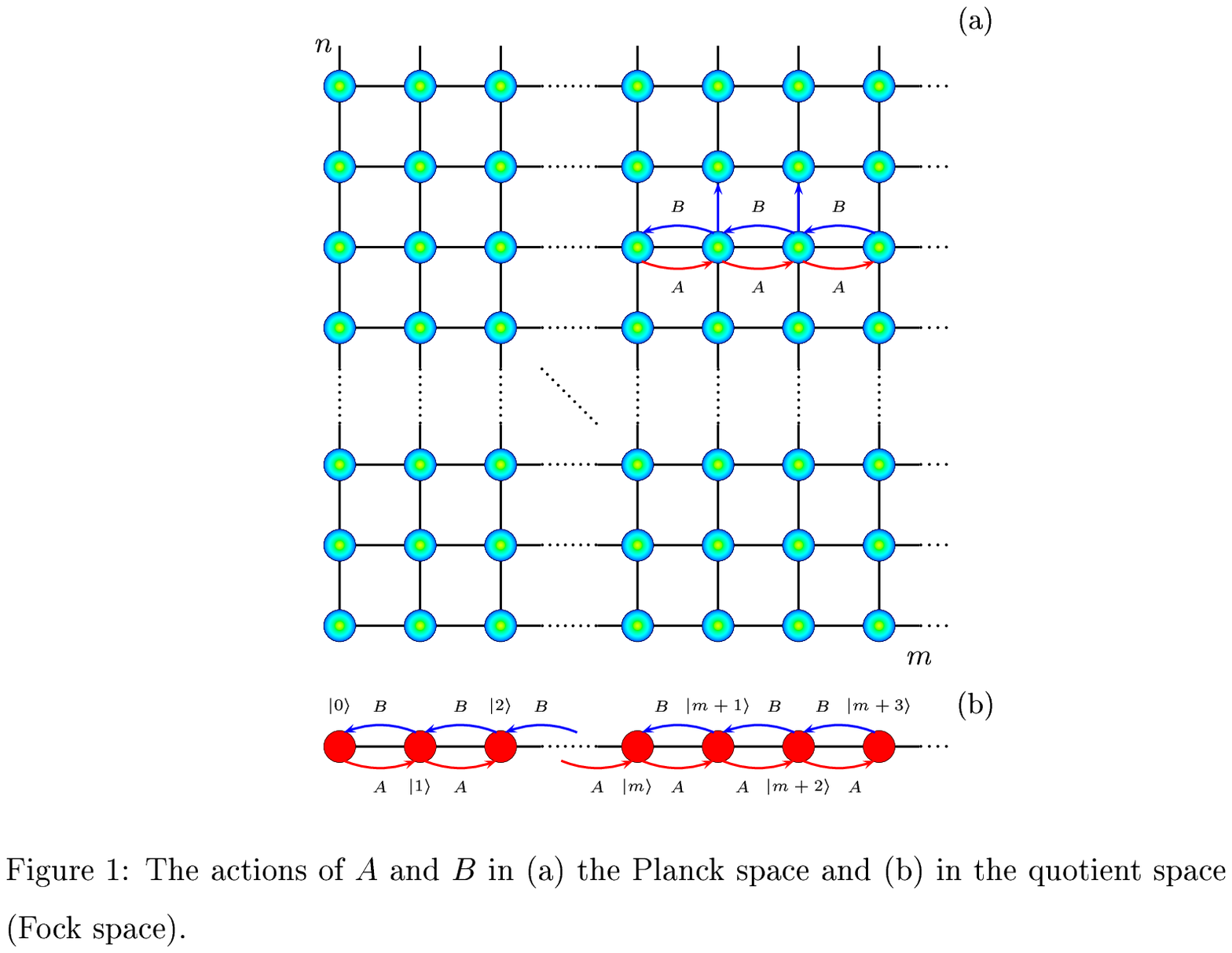}
\end{figure}
This representation of Heisenberg-Weyl algebra is indecomposable and can be
applied to construct new type representations of Lie (super) algebras,
Kac-Moody algebras and quantum groups [4]. We observe that
\[
V_1=span\{f(m,1+n)|m,n\in Z^{+}\}
\]
is a $\mathcal{W}_1-$invariant subspace. In Fig.1, the equivalence classes
are just the vertical lines and the corresponding quotient space
\[
V/V_1=span\{\left| m\right\rangle \triangleq \frac{f(m,0)}{\sqrt{m!}}%
ModV_1|m\in Z^{+}\}.
\]
is represented as a horizontal line $(m,0)(m=0,1,\cdots )$.Thus $\mathcal{W}%
_1$ has an induced representation on $V/V_1$,as is illustrated in Fig1.b.
This is just the so called Fock representation. Indeed, with the induced
dynamic operators $a^{\dagger }=\overline{A}$ and $a=\overline{B}$ we have
\begin{eqnarray}
a^{\dagger }\left| m\right\rangle  &=&\sqrt{m+1}\left| m+1\right\rangle  \\
a\left| m\right\rangle  &=&\sqrt{m}\left| m-1\right\rangle   \nonumber
\end{eqnarray}

Now suppose an evolution process at the Planck scale is described by the
``Hamiltonian''
\begin{equation}
\mathcal{H}=AB\in End(V).
\end{equation}
If we define the inner product $(\ ,\ )$ on $V$ as $(f(m,n),f(m^{\prime
},n^{\prime }))=\delta _{mm^{\prime }}\delta _{nn^{\prime }}$,$\mathcal{H}$
is not Hermitian. Then $\mathcal{H}$ may interpreted as describing a
dissipative process. On the other hand,if we introduce an inner product on $%
V/V_1$ such that $\left\langle m|n\right\rangle =\delta _{mn},$the induced
operator
\begin{equation}
\overline{\mathcal{H}}=a^{\dagger }a\in End(V/V_1)
\end{equation}
is just Hermitian. Hence if we identify $V/V_1$ with the space of quantum
states, then $\overline{\mathcal{H}}$ describes a unitary quantum mechanical
evolution. In fact, it is the Hamiltonian of a quantum oscillator. Although
this overly simplistic example is not convincing enough in real physics, it
nevertheless suggests that generalization of 't Hooft's idea might be
fruitful.

\section{Remarks}

To conclude this paper we should give some remarks on our mathematical
reformulation and the physical generalization for 't Hooft's equivalence
class theory . Firstly, a correct quantum theory requires a Hilbert space
with properly defined inner product to define probability. But it is not at
all clear how to endow the space of equivalence classes with such a inner
product even though there may be a natural inner product on the space of
primordial states. Thus to establish the unitarity of the induced evolution
is really a problem if one does not know in advance what the physical system
at the atomic scale seems to be . So a gap remains to be bridged between the
so called Planck scale physics and the atomic scale physics even if 't
Hooft's theory proves to be correct. Secondly, it is challenging to
understand quantum decoherence or wave function collapse in quantum
measurement [5]from the underlying deterministic theory at a deeper level.
However, like the hidden variable theory , which has been rejected by
experiments till now, quantum measurement problems such as quantum-classical
correspondence, quantum dissipation and quantum entanglement[5,6] ) must be
faced if we are to take 't Hooft's theory seriously. \vspace{1cm}

\noindent \textit{\ This work is supported by the NFS of China. One of the
authors (CPS) would like to express his sincere thanks to Professor Stephen
Adler for a useful discussion about the t'Hooft 's work with him.}

\end{document}